\documentclass[acmsmall,screen,nonacm]{acmart}
\def\BibTeX{{\rm B\kern-.05em{\sc i\kern-.025em b}\kern-.08emT\kern-.1667em\lower.7ex\hbox{E}\kern-.125emX}}
    
\usepackage{silence}
\ActivateWarningFilters[todo]
\usepackage{multirow}

\usepackage{pifont}

\newcommand{\cmark}{\ding{51}}%
\newcommand{\xmark}{\ding{55}}%
\begin{document}

%
\title{Mapi-Pro: An Energy Efficient Memory Mapping Technique for Intermittent Computing}

%
\author{SatyaJaswanth Badri}
\email{2018CSZ0002@iitrpr.ac.in}
\author{Mukesh Saini}
\email{mukesh@iitrpr.ac.in}

\author{Neeraj Goel}
\email{neeraj@iitrpr.ac.in}

\affiliation{%
  \institution{Indian Institute of Technology Ropar}
  \streetaddress{S.Ramanujan Block, IIT Ropar Main Campus}
  \city{Ropar}
  \state{Punjab, India}
  \postcode{140001}
}

%
\renewcommand{\shortauthors}{S.J Badri, et al.}

%
\begin{abstract}
Battery-less technology evolved to replace battery usage in space, deep mines, and other environments to reduce cost and pollution. Non-volatile memory (NVM) based processors were explored for saving the system state during a power failure. Such devices have a small SRAM and large non-volatile memory. To make the system energy efficient, we need to use SRAM efficiently. So we must select some portions of the application and map them to either SRAM or FRAM. This paper proposes an ILP-based memory mapping technique for Intermittently powered IoT devices. Our proposed technique gives an optimal mapping choice that reduces the system's Energy-Delay Product (EDP). We validated our system using a TI-based MSP430FR6989 and MSP430F5529 development boards. Our proposed memory configuration consumes 38.10\% less EDP than the baseline configuration and 9.30\% less EDP than the existing work under stable power. Our proposed configuration achieves 15.97\% less EDP than the baseline configuration and 21.99\% less EDP than the existing work under unstable power. This work supports intermittent computing and works efficiently during frequent power failures. 

\end{abstract}

%
%

%

%

%
\maketitle

\section{Introduction}

The Internet of Things (IoT) is a network of sensors and nodes that allows nearby objects to communicate and collaborate easily. Batteries are the most common source of power for IoT devices. Because of the battery's limited capacity and short lifespan \cite{lifetime}, replacement is costly. IoT may consist of billions of sensors and systems by the end of 2050 \cite{big}. Replacing and disposing billions of battery-operated devices is expensive and hazardous to the environment. As a result, we need battery-free IoT devices.

Energy harvesters are a promising alternative to battery-powered devices. The energy harvester collects energy from the environment and stores energy in capacitors. Energy harvesting is unreliable, power failures are unavoidable, and the application's execution is irregular. This type of computing is known as intermittent computing \cite{intermittent,int2,int3}. 

For intermittently powered IoT devices, energy harvesting is the primary energy source. Energy harvesting sources like piezo-electric materials and radio-frequency devices extract a small amount of energy from their surroundings. We must use energy efficiently in both stable and unstable power supply scenarios.

In order to utilize energy efficiently and to make the system energy efficient, we primarily have two choices. The first choice is to reduce energy consumption by proposing new techniques that use energy efficiently. The second choice is to increase the number of different energy harvesters, which will accumulate more energy while increasing maintenance costs. We need to maintain these many energy harvesters, which is not a feasible solution. Thus, our main concern is to reduce energy consumption by proposing new techniques which help to design an energy-efficient system.

Gonzalez et al. \cite{edp} mentioned energy as not an ideal metric for evaluating system efficiency. By simply reducing supply voltage or load capacitance, energy can be reduced. Instead of using energy as a metric, they suggested using the Energy-Delay Product (EDP) as the energy-efficient design metric. The EDP considers both performance and energy simultaneously in a design. If a design minimizes the EDP, we can call such a design energy-efficient. We define EDP in the equation \ref{eq1}.

\begin{equation}\label{eq1}
      E D P= E_{system} \times Num\_cycles
\end{equation}

Where $E_{system}$ is the system's energy consumption, $Num\_cycles$ is the number of CPU cycles.


During these frequent power failures, executing IoT applications becomes more difficult because all computed data may be lost, and the application's execution must restart from the beginning. During power failures, we need an additional procedure to backup/checkpoint the volatile memory contents to non-volatile memory (NVM).

Flash memory was the prior NVM technology used by modern microcontrollers at the main memory level, such as MSP430F5529 \cite{msp1}. Flash is ineffective for frequent backups and checkpointing because it's erase/write operations require a lot of energy. Emerging NVMs outperform flash, including spin-transfer-torque RAM (STT-RAM) \cite{c3,stt1}, phase-change memory (PCM) \cite{c8}, resistive RAM (ReRAM), and ferroelectric RAM (FRAM) \cite{msp}. Previous works have been demonstrated by incorporating these emerging NVMs into low-power-based microcontrollers (MCUs)  \cite{msp,msp1,r3}. Recent NVM-based MCUs, such as the flash-based MSP430F5529 and the FRAM-based MSP430FR6989, encourage the use of hybrid main memory. The flash-based MCU, MSP430F5529, is made up of SRAM and flash, while the FRAM-based MCU, MSP430FR6989, is made up of SRAM and FRAM at the main memory level. The challenges associated with hybrid main memory-based architectures, such as MSP430FR6989, are as follows.


\begin{enumerate}
 \item FRAM consumes 2x times more energy and latency than SRAM. This design degrades system performance and consumes extra energy even during normal operations. 

\item SRAM loses contents during a power failure and needs to execute the application from the beginning, which consumes extra energy and time. For large-size applications, this design will not be helpful. Anyway, using only SRAM performs better during regular operations.

\item We can design a hybrid main memory to get the benefits from both SRAM and FRAM. The following questions need to be answered and analyzed to use the hybrid main memory design.

\begin{enumerate}
    \item How do we choose the appropriate sections of a program and map them to either SRAM or FRAM regions? A significant challenge is mapping a program's stack, code, and data sections to either SRAM or FRAM.
\item  How and where should volatile contents be backed up to the NVM region during frequent power failures?
\end{enumerate}

\end{enumerate}



The main question is which section of an application should be placed in which memory region; this is essentially a memory mapping problem. Concerning all of the challenges mentioned earlier, this article makes the following contributions:


\begin{itemize}
    \item To the best of our knowledge, this is the first work on the Integer-Linear Programming (ILP) based memory mapping technique for intermittently powered IoT devices. 
    
    \item We incorporated the energy-harvesting scenarios into the ILP model such that the frequency of power failures is considered as an input for our ILP model.

\item We formulated the memory mapping problem to cover all the possible design choices. We also formulated our problem in such a way that it supports large-size applications.

\item We proposed a framework that efficiently consumes low energy during regular operation and frequent power failures. Our proposed framework supports intermittent computing.

\item We evaluated the proposed techniques and frameworks in actual hardware boards.
\end{itemize}

Our proposed ILP model recommends placing each section in either SRAM or FRAM. We compared the proposed memory configuration and techniques with the baseline memory configurations under both stable and unstable power scenarios. Our proposed memory configuration consumes 38.10\% less EDP than the FRAM-only configuration and 9.30\% less EDP than the existing work under stable power. Our proposed configuration achieves 15.97\% less EDP than the FRAM-only configuration and 21.99\% less EDP than the existing work under unstable power.

\textbf{Paper organization:} Section \ref{p2} discusses the background and related works. Section \ref{moti} explains the motivation behind the proposed framework. Section \ref{system1} explains the system model and gives an overview of the problem definition. Section \ref{p4} explains about proposed ILP-based memory mapping technique and framework that supports during frequent power failures. The experimental setup and results are described in section \ref{exp1}. We conclude this work in section \ref{p6}.

\section{Background and Related works} \label{p2}

SRAM and DRAM are used to design registers, caches, and main memory in traditional processors. For an intermittently aware design, we replace a regular processor's volatile memory model with an NVM. STT-RAM, PCM, flash, and FRAM are all relatively new NVM technologies \cite{c1,c3,c4,c6,c8,xie,pcm,stt1,stt2,choi}. FRAM consumes less energy than other NVM technologies, such as Flash. FRAM can be helpful for IoT devices that are operating at low power. These NVM technologies motivated researchers because of their appealing characteristics, such as non-volatility, low cost, and high density \cite{m6,m7,m8,cache1n,cache2n}. 

Researchers started using real-time NVM-based MCUs for intermittent computing \cite{nvp1,nvp2,msp,msp1,m9}. Researchers observed that using only NVMs at the cache or main memory level degrades the system's performance and consumes more energy, which gives an idea to explore hybrid memories. Recent NVM-based MCU such as MSP430FR6989 \cite{msp} consists of both SRAM and FRAM. We need to utilize the SRAM and FRAM efficiently and correctly; otherwise, we may degrade system performance and consume extra energy. To make the system more efficient, we need to map the application contents to either SRAM or FRAM. This is actually a memory mapping problem, similar to scratch-pad memories.

Recent works focused on incorporating NVMs as a virtual memory during frequent power failures. Andrea et al. \cite{m1} propose Alfred, a mapping technique that maps virtual memory to volatile or NVM. Andrea et al. use machine-level codes for these mappings that achieve 2x improvement compared to existing techniques and systems. However, the Andrea et al. technique doesn't discuss the complete design choices or consider the real-time power scenarios.  

Including NVMs in systems need to answer the following research questions: when to a checkpoint and where to checkpoint the volatile data. Researchers proposed efficient checkpointing techniques \cite{m2,m5} incorporating user-defined function calls that help determine how much energy is still available in the capacitor. Based on that analysis, the system invokes the checkpoint. These techniques even predict power failures and support intermittent computing. 

Researchers explored a similar mapping problem in scratch-pad memories (SPMs) \cite{r1_n, r2_n, r3_n}. Chakraborty et al. \cite{partitioning} documented the existing and standard memory mapping techniques on SPMs. In earlier works, memory mapping was done mainly between SPMs and main memory. Memory mapping can be done statically and dynamically \cite{d1,d2}. In static memory mapping, either ILP or the compiler can assist in determining the best placement \cite{r1_n, r2_n, m4, r3_n}. ILP-solver takes inputs obtained from profilers and memory sizes as constraints in ILP-based memory mapping works. The ILP-solver provides the best placement option based on the objective function. In dynamic memory allocation \cite{d11,d21,d31,d41}, either the user-defined program or the compiler will decide on an optimal placement choice at run time.

However, our problem differs from the memory mapping techniques in SPMs because intermittent computing brings new constraints. During intermittent computation, the challenges were the forward progress of an application, data consistency, environmental consistency, and concurrency between the tasks. Due to these challenges, the execution model and development environment differ from the SPM-based memory mapping techniques. As a result, we require a memory mapping technique that supports intermittent computation.

Researchers have explored memory mapping techniques and analysis for the MSP430FR6989 MCU. Kasım et al. \cite{m3} proposed a task-based mapping mechanism considering all event-driven paradigms that support intermittent computing and battery-less sensing devices. In FRAM-based MCUs, Jayakumar et al. \cite{r3} implement a checkpointing policy. They save the system state to FRAM during a power failure. Jayakumar et al. \cite{r4, backup} propose an energy-efficient memory mapping technique for TI-based applications in FRAM-based MCUs. Kim et al. \cite{r5} presents a detailed analysis of energy consumption for all memory sections in FRAM-based MCUs under different memory mappings.

Earlier works investigated this problem by analyzing the possibilities to make the system efficient. The authors \cite{r4, backup, r5} have not covered all the design choices and possibilities. In addition, there is significantly less contribution towards memory mappings in FRAM-based MCUs that supports intermittent computation. Our work proposes an energy-efficient memory mapping technique for intermittently powered IoT devices that experience frequent power failures.

\section{Motivation} \label{moti}

This section discusses the advantages of using hybrid SRAM and FRAM for these MSP430-based MCUs over SRAM-only or FRAM-only designs, as well as the importance of efficient memory allocation. 

\begin{figure}[htbp]
  \includegraphics[width= 1\linewidth]{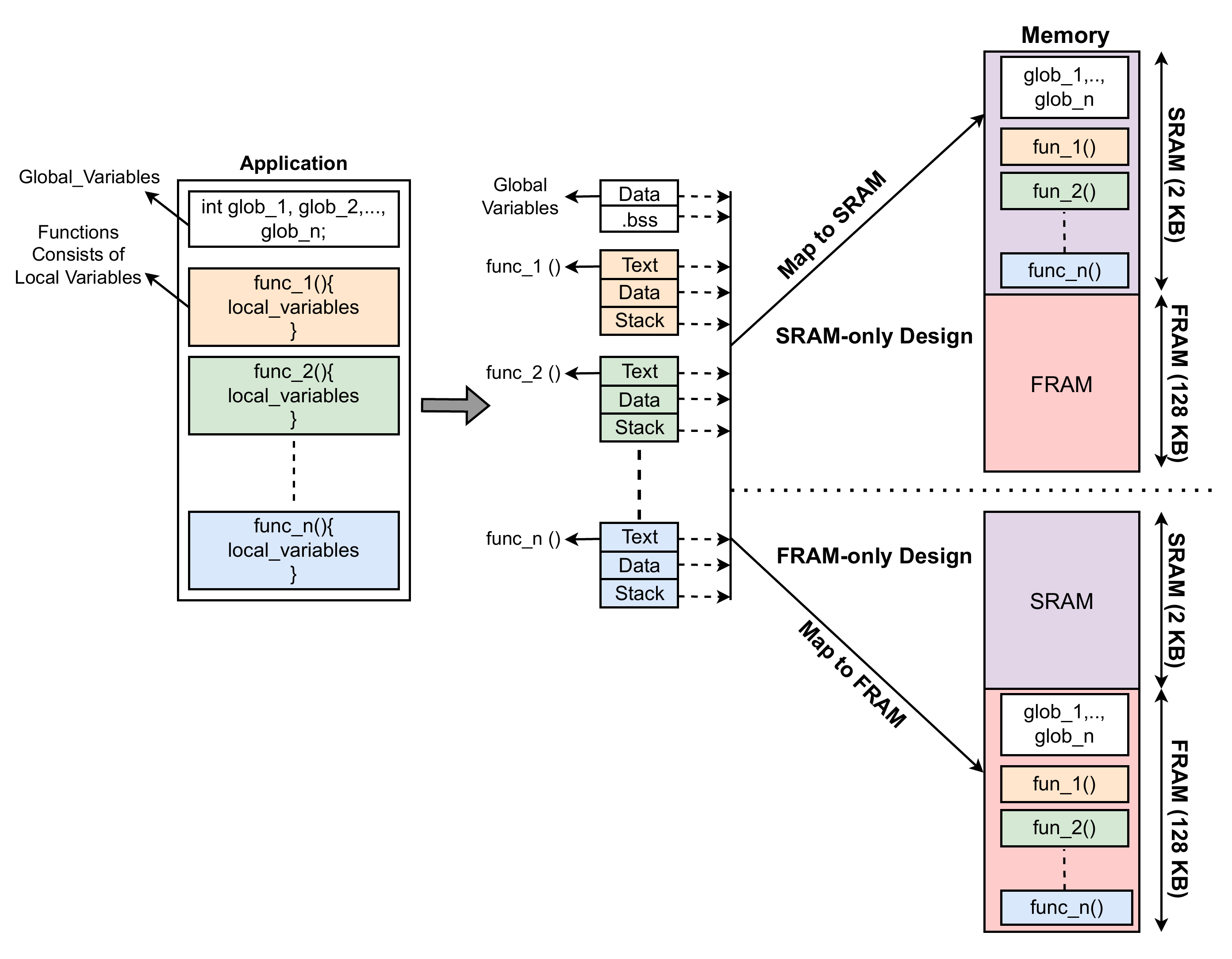}
\caption{Overview of the FRAM-only configuration and SRAM-only configuration memory mappings in MSP430FR6989}
\label{b11}
\end{figure}

SRAM is 2KB, and FRAM is 128KB in a FRAM-based MCU, MSP430FR6989. The first naive approach is to use the entire 128KB of FRAM in both stable and unstable power scenarios, resulting in longer execution cycles and higher energy consumption. Similarly, we have a second naive approach to use the entire 2KB SRAM for small applications (whichever fits within the SRAM size), which has advantages during regular operation. Unfortunately, it loses all 2KB SRAM data during a power failure and takes more time to backup 2KB contents to FRAM during a power failure. As shown in figure \ref{b11}, for the FRAM-only configuration, we map all three sections to FRAM and all three sections to SRAM for the SRAM-only configuration. 

\begin{figure}[htpb]
\centering
\begin{minipage}{.5\textwidth}
  \centering
  \includegraphics[width=\linewidth]{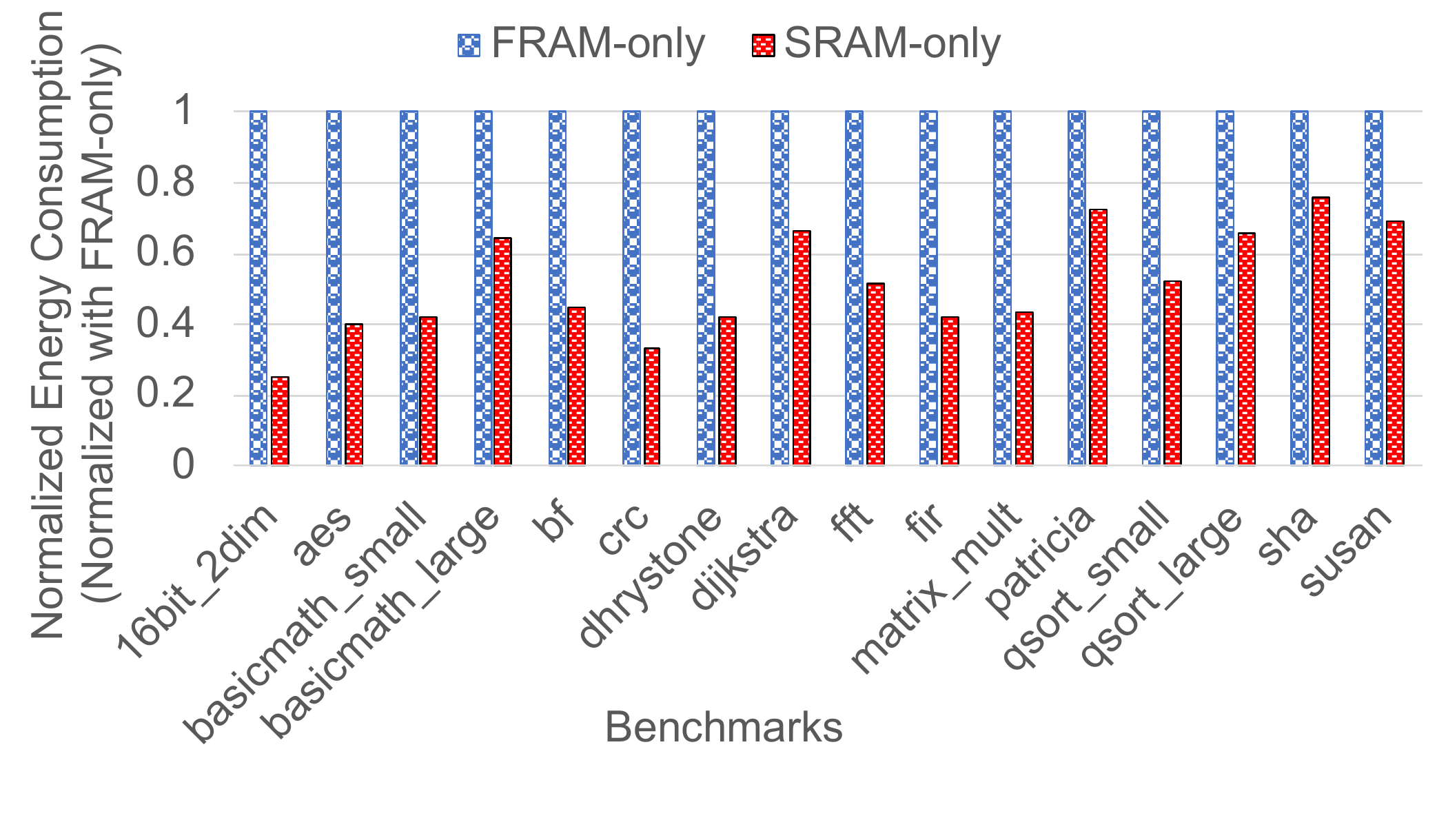}
     \centerline{(a) Under Stable Power }\medskip
\end{minipage}%
\begin{minipage}{.5\textwidth}
  \centering
  \includegraphics[width=\linewidth]{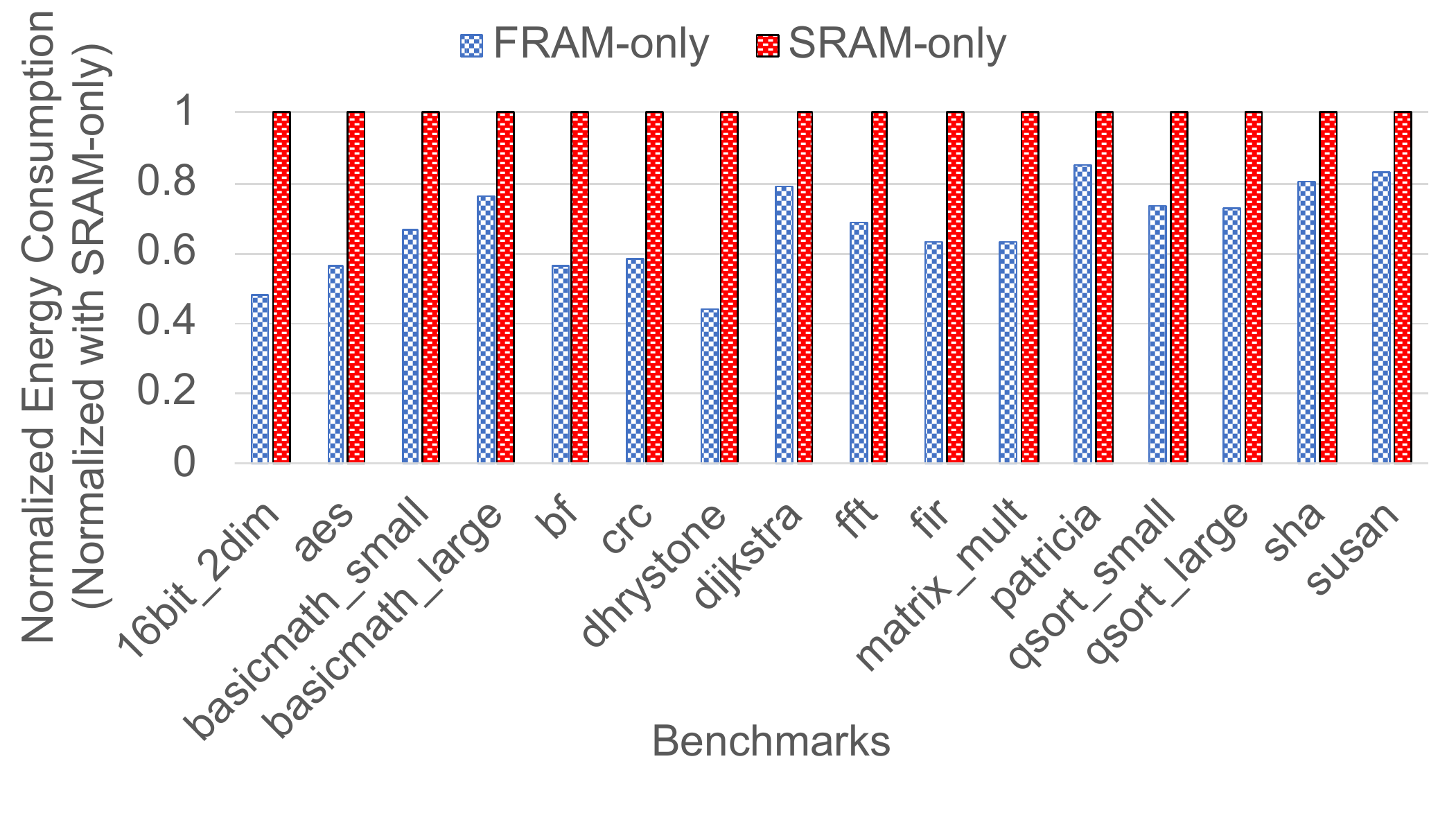}
  \centerline{(b) Under Unstable Power }\medskip
\end{minipage}

\hfill
\caption{Comparison between FRAM-only and SRAM-only configurations under Stable and Unstable Power}
\label{figt3}
\end{figure}

We compared FRAM-only configuration and SRAM-only configuration in both stable and unstable power scenarios. FRAM-only configuration performs better during frequent power failures, while SRAM-only configuration performs better during regular operations (without any power failures), as shown in figure \ref{figt3}. On average, FRAM-only configuration consumes 47.9\% more energy than SRAM-only configuration during a stable power, as shown in figure \ref{figt3} (a). On average, SRAM-only configuration consumes 32.7\% more energy than FRAM-only configuration during an unstable power, as shown in figure \ref{figt3} (b). We also observed that MCU would pitch an error to either increase the SRAM space or use FRAM space for any computations. For large-size applications will not run using only SRAM, it requires FRAM as well. Thus, large applications consume more energy in SRAM-only configuration during a stable power scenario.

These two designs motivate us to propose a hybrid memory design that effectively uses both SRAM and FRAM. We also encountered that SRAM-only configuration is ineffective for larger applications. As a result, we had to use a hybrid memory and figure out how and where to place the sections. To the best of our knowledge, only one work explored the memory mapping issue for these MCUs \cite{backup}. We analyzed the mapping decisions using their empirical model. Jayakumar et al. \cite{backup} calculated the energy consumption values for each configuration. The authors suggested that allocate the sections to either SRAM or FRAM based on the energy values.

The empirical method used by the authors is as follows. The authors considered functions as the basic unit. They explored all configurations and calculated the energy values, as shown in table \ref{emp}. The authors have eight configurations because they have two memory regions (SRAM or FRAM) and need to map three sections (stack, data, text). Using the author's model, we calculated the energy values for the qsort\_small application. For instance, the \{SSS\} configuration performs better during a stable power supply, and during a power failure, \{SFS\} consumes less energy than all other configurations. As a result, authors allocate text and stack sections to SRAM and data sections to FRAM. 

\begin{table}[htp]
\centering
\caption{Analysis of the Empirical Methods Used by Jayakumar et al. \cite{backup} for qsort\_small Under Stable and Unstable Power Scenarios}
\label{emp}
\begin{tabular}{|l|l|l|l|l|l|}
\hline
\textbf{Configuration} & \textbf{Text} & \textbf{Data} & \textbf{Stack} & \textbf{$Energy_{stable} (mJ)$} & \textbf{$Energy_{unstable} (mJ)$} \\ \hline
1. \{SSS\}                  & SRAM                  & SRAM                  & SRAM                   &    16.70   &     79.56     \\ \hline
2. \{SSF\}                   & SRAM                  & SRAM                  & FRAM                   &    21.08    &    66.34     \\ \hline
3. \{SFS\}                   & SRAM                  & FRAM                  & SRAM                   &     28.75    &   33.79     \\ \hline
4.   \{SFF\}                 & SRAM                  & FRAM                  & FRAM                   &    35.97     &    52.10    \\ \hline
5.   \{FSS\}                 & FRAM                  & SRAM                  & SRAM                   &    39.48      &   68.24    \\ \hline
6.  \{FSF\}                  & FRAM                  & SRAM                  & FRAM                   &     57.64     &    54.75   \\ \hline
7.   \{FFS\}                 & FRAM                  & FRAM                  & SRAM                   &      64.14    &  45.83     \\ \hline
8.    \{FFF\}                & FRAM                  & FRAM                  & FRAM                   &       92.09    &   36.07   \\ \hline
\end{tabular}%
\end{table}

We observed that this empirical method becomes ineffective as the number of configurations increases. The authors considered all global variables, arrays, and constants as data sections. Instead, why can't we map each global variable or array to either SRAM or FRAM? This increases the number of configurations, and calculating/tracking energy values is challenging. Our design space grows enormously and makes our mapping problem challenging.

This new set of challenges motivated us to propose an energy-efficient memory mapping technique. Our proposed memory mapping framework supports large-size applications and covers all possible configurations.

\section{System model and Problem Definition} \label{system1}
This section discusses the system model for embedded MCUs and defines the mapping problem for these MCUs.

\subsection{System Model} \label{mot}
We consider a simple, customized RISC instruction set with a Von-Neumann architecture, where the instructions and data share the same address space that supports at least 16-bit addressing. Base architecture doesn't have a cache to avoid uncertainty. To make things simple, we assume single cycle execution of the processor. Base architecture has a small SRAM memory and a larger NVM.

The MSP430 is an example of such a processor. Non-volatile memory sizes range from 1 kilobyte (KB) to 256 KB, while volatile RAM sizes range from 256 bytes to 2KB. Both SRAM and NVM can be accessed by instructions using a compiler/linker script. We can modify the linker script to map memory according to the memory ranges specified by the user. MSP430 doesn't have any operating system.

\subsection{Problem Definition} \label{p3}
\textit{Definition 4.1:} \textbf{Optimal Memory Mapping Problem}: Given a program that consists of various functions and global variables, sizes of SRAM and FRAM, the number of reads and writes for each function/variable, frequency and duration of power failures, and the energy required per read/write to the SRAM/FRAM. What is the optimal memory mapping for these functions/variables in order to reduce the system's EDP?

\textbf{The inputs are :} The number of functions; the number of global variables; energy per write to SRAM and FRAM; energy per read to SRAM and FRAM; SRAM and FRAM sizes; Number of CPU cycles per each function; the number of reads; the number of writes; frequency and duration of power failures.

\textbf{The output is:} Mapping information for all functions and global variables, under which the system's EDP is minimized.

\textit{Definition 4.2:} \textbf{Support for Intermittent Computing}: During power failures, we must safely backup the volatile contents to NVM. As previously stated, we must use SRAM efficiently for energy savings; but again, how can we save the contents of SRAM? There are two significant issues with intermittent computation. First, during a power failure, all SRAM's mapping information and register contents are lost, causing the system to become inconsistent. Second, how do we backup/restore the mapping information and register contents to ensure system consistency?

Our first objective is to minimize the overall system's energy and system's EDP. We need to support the proposed system even during frequent power failures. Our second objective is to maximize the execution progress of the application during frequent power failures. Application progress is the function of both execution time and the frequency of power failures ($\eta$), as shown in equation \ref{eq1100}. 

\begin{equation} \label{eq1100}
\text { Progress }=F \text { ($NC_{Execute} * \eta$) }
\end{equation}

We define the frequency of power failures in equation \ref{eq110}. It is the ratio between the time consumed during regular execution without any power failures to the time consumed during power failures, where $NC_{Execute}$ is the number of cycles required for executing the application during regular operation.

\begin{equation} \label{eq110}
\begin{split}
        \eta &= \frac{NC_{Execute}}{NC_{Intermittent}}
\end{split}
\end{equation}

We define the number of cycles required during power failures in equation \ref{eq1121}. We need to perform the backup and restore operations during a power failure. 

\begin{equation} \label{eq1121}
NC_{Intermittent} =  NC_{Backup} + NC_{Execute} + NC_{Restore}
\end{equation}

Where $NC_{Backup}$ is the number of cycles required for the backup operation, and $NC_{Restore}$ is the number of cycles required for the restore operation.

\section{Mapi-Pro: An Energy Efficient Memory Mapping for Intermittent Computing} \label{p4}

\begin{figure*}[htbp]
  \includegraphics[width= 1\linewidth]{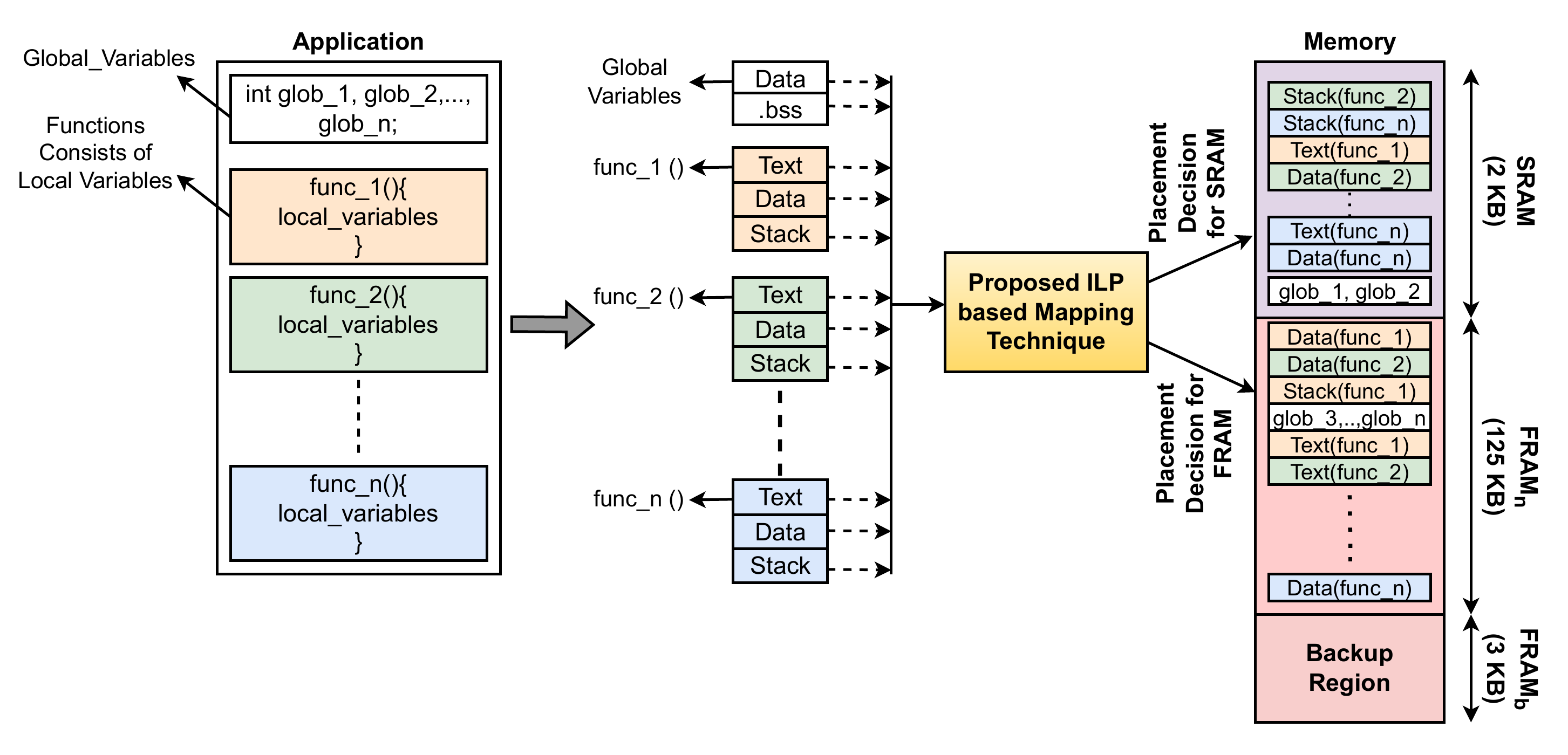}
\caption{Overview of the proposed memory mappings in MSP430FR6989}
\label{md1}
\end{figure*}

In this section, we discuss the details of the proposed mapping technique. Our main objective is to pick the optimal mapping choice among all the design choices, which reduces the system's EDP. To achieve this, we proposed an ILP-based mapping technique. The overview of the proposed mapping technique is shown in figure \ref{md1}. We also discuss how we support intermittent computing for these MCUs.  

\subsection{ILP Formulation for Data Mapping for Intermittent Computing}
\begin{figure*}[htbp]
  \includegraphics[width= 1\linewidth]{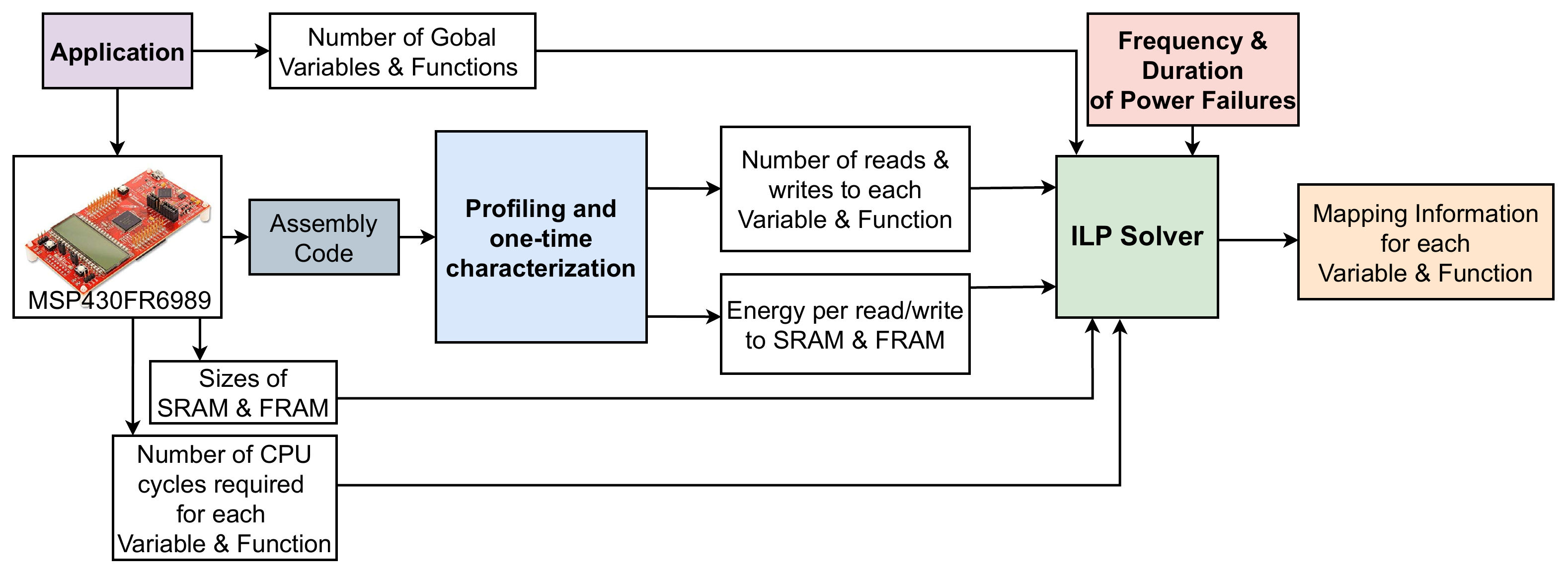}
\caption{Overview of the Proposed ILP Framework}
\label{f1}
\end{figure*}

We present the ILP formulation for the memory mapping problem mentioned in definition 4.1. We divide this ILP formulation into two parts, one is for global variables, and the second is for the functions. We have shown the overview block diagram of the proposed ILP framework in figure \ref{f1}. During the profiling and characterization process, we consider the branch instructions and their behavior from the generated assembly code.

\textbf{For Global Variables:} Let the number of global variables in a program be `G'. Let the number of reads and writes to variable `i' are $r_i$ and $w_i$. We divided FRAM's 128 KB into two regions, i.e., $FRAM_n$ and $FRAM_b$, $FRAM_n$ memory region has 125 KB, and the $FRAM_b$ memory region has 3 KB. 

We have two memory regions represented as $Mem_j$ as shown in the equation \ref{eq2.1}; when j=1, we select the memory region as SRAM, and we use $FRAM_{n}$ for j=2.

\begin{equation} \label{eq2.1}
    Mem_j = \begin{cases}j = 1 & \text {; SRAM } \\ j = 2 & ; FRAM_{n} \end{cases}
\end{equation}

Let the sizes of SRAM/FRAM as $Size(Mem_j)$ as shown in equation \ref{eq2}, when j=1, we refer as SRAM memory size in bytes, and when j=2, we refer as $FRAM_{n}$ memory size in bytes. 

\begin{equation} \label{eq2}
    Size(Mem_j) = \begin{cases}j = 1 & \text {; SRAM } \\ j = 2 & ; FRAM_{n} \end{cases}
\end{equation}

Let the energy required for each read/write to $Mem_j$ is $E_{r\_j}$ and $E_{w\_j}$. Let the number of CPU cycles required to execute a global variable $v_i$ be $NC_{v_i}$, where $\forall i \in[1, G])$. Using one-time characterization and static profiling, we gathered data such as per read/write energy to SRAM/FRAM and the number of cycles.

We define a binary variable (BV); $I_{j}\left(v_{i}\right)$, which refers to a variable $v_i$ is allocated to memory region $j$. If $I_{j}\left(v_{i}\right)$=1 then the variable $v_i$ is allocated and $I_{j}\left(v_{i}\right)$=0 indicates that the variable $v_i$ is not allocated. $I_{j}\left(v_{i}\right)$, where $(\forall j \in[1, Mem_j], \forall i \in[1, G])$ is defined as shown in the equation \ref{eq5}. 

\begin{equation} \label{eq5}
I_{j}\left(v_{i}\right)= \begin{cases}1 & v_{i} \text { is allocated to memory region } j \\ 0 & \text { otherwise}\end{cases}\end{equation}

\textbf{Constraints:} There are two constraints, one is for BV; $I_{j}\left(v_{i}\right)$ and one is a memory size constraint. In any case, a variable $v_i$ is allocated to only one memory region, which means $v_i$ is allocated to either SRAM or FRAM but not both. This constraint is defined in the equation \ref{eq6}.

\begin{equation}  \label{eq6}
\sum_{j=1}^{Mem_j} I_{j}\left(v_{i}\right)=1 \quad(\forall i \in[1, G])
\end{equation}

The other constraint is related to memory sizes. The allocated variables $v_i$ and its $Size(v_i)$; $\forall i \in[1, G])$ should not be greater than the $Size(Mem_j)$. This constraint is defined in the equation \ref{eq7}.

\begin{equation}  \label{eq7}
\sum_{i=1}^{G} I_{j}\left(v_{i}\right) * Size(v_i) \leq Size(Mem_j) \quad(\forall j \in[1, Mem_j])
\end{equation}

\textbf{Objective 4.1: } 
The challenge of mapping global variables in a program to either SRAM or FRAM is to reduce EDP and improve system performance. $E_{global}$ is defined in the equation \ref{eq8}. Where $E_{global}$ is the energy required to allocate global variables to either SRAM or FRAM and execute those from their respective memory regions.

\begin{equation} \label{eq8}
E_{global} = \sum_{j=1}^{Mem_j} \sum_{i=1}^{G} [E_{r\_j} \times r_{i} +E_{w\_j} \times w_{i}] 
\end{equation}

$EDP_{global}$ is defined in the equation \ref{eq911}. Where $EDP_{global}$ is the energy-delay product required to allocate global variables to either SRAM or FRAM.

\begin{equation} \label{eq911}
EDP_{global} = \sum_{j=1}^{Mem_j} \sum_{i=1}^{G} I_{j}\left(v_{i}\right) [ E_{global} \times NC_{v_i} ]
\end{equation}

\textbf{For Functions: } Let the number of functions in a program be $`N_f'$. Let the number of reads and writes to $i^{th}$ function are $r(F_i)$ and $w(F_i)$, where $\forall i \in[1, N_f]$. Functions consist of procedural parameters, local variables, and return variables. Internally the code/data of functions are divided into the text, data, and stack sections. We map at least one section among these three sections to either SRAM or FRAM regions, i.e., $Mem_j$ and $Sec_k(i)$ defines section `k' of $i^{th}$ function as shown in the equation \ref{eq10}, when k=1, we refer to the text section of $i^{th}$ function, when k=2, we refer to the data section of $i^{th}$ function, and when k=3, we refer to the stack section of $i^{th}$ function.

\begin{equation} \label{eq10}
    Sec_k(i) = \begin{cases}k = 1 & \text {; Text } \\ k = 2 &  \text {; Data } \\ k = 3 &  \text {; Stack } \end{cases} ; \forall i \in[1, N_f]
\end{equation}

We define a BV; $I_{j}\left(Sec_{k}(i) \right)$, which refers to a section $Sec_k$ of $i^{th}$ function is allocated to only one memory region $j$. If $I_{j}\left(Sec_{k}(i) \right)$=1 then the section $Sec_i$ is allocated and $I_{j}\left(Sec_{k}(i) \right)$=0 that indicates the section $Sec_i$ is not allocated. $I_{j}\left(Sec_{k}(i) \right)$, where $(\forall j \in[1, Mem_j], \forall i \in[1, N_f])$, $\forall k \in[1, Sec_k(i)])$ is defined as shown in the equation \ref{eq11}. 

\begin{equation} \label{eq11}
I_{j}\left(Sec_{k}(i) \right) = \begin{cases}1 & Sec_k \text{ of }  i^{th} \text { function is allocated to } j \\ 0 & \text { otherwise}\end{cases} \end{equation}

\textbf{Constraints:} There are two constraints, one is for BV; $I_{j}\left(Sec_{k}(i) \right)$ and one is a memory size constraint. In any case, a $Sec_k$ of $i^{th}$ function is allocated to only one memory region, which means $Sec_k$ of $i^{th}$ function is either allocated to either SRAM or FRAM but not both. This constraint is defined in the equation \ref{eq12}.

\begin{equation}  \label{eq12}
\sum_{k=1}^{3} \sum_{j=1}^{Mem_j} I_{j}\left(Sec_{k}(i) \right))=1 \quad(\forall i \in[1, N_f])
\end{equation}

The other constraint is related to memory sizes. The allocated sections $Sec_{k}(i)$ and its $Size(F_i)$; $\forall k \in[1, Sec_k(i)])$, $\forall j \in[1, Mem_j]$, $\forall i \in[1, N_f] $ should not be greater than the $Size(Mem_j)$. This constraint is defined in the equation \ref{eq13}.

\begin{equation}  \label{eq13}
\sum_{i=1}^{G} I_{j}\left(v_{i}\right) * Size(v_i) + \sum_{k=1}^{3} \sum_{i=1}^{N_f} I_{j} \left(Sec_{k}(i) \right) * Size(F_i) \leq Size(Mem_j)
\end{equation}


\textbf{Objective 4.2: } 
The challenge of mapping sections of these functions in a program to either SRAM or FRAM is to minimize EDP and improve system performance. $E_{func}$ is defined in the equation \ref{eq14}, where $M_{c_i}$ is the number of the times $i^{th}$ functions called. 
\begin{equation} \label{eq14}
E_{func} = \sum_{j=1}^{Mem_j} \sum_{i=1}^{N_f} [E_{r\_j} \times r(F_i) +E_{w\_j} \times w(F_i)] \times M_{c_i}
\end{equation}

$EDP_{func}$ is defined in the equation \ref{eq912}. Where $EDP_{func}$ is the energy-delay product required to allocate all functions to either SRAM or FRAM. Where $E_{func}$ is the energy required to allocate functions to either SRAM or FRAM. Where $NC_{F_i}$ is the number of CPU cycles required to execute a function $F_i$.

\begin{equation} \label{eq912}
EDP_{func} = \sum_{k=1}^{3} \sum_{j=1}^{Mem_j} \sum_{i=1}^{N_f} I_{j} \left(Sec_{k}(i) \right) [ E_{func} \times NC_{F_i} ]
\end{equation}

The overall system EDP, $EDP_{system}$, is the sum of both $EDP_{global}$ and $EDP_{func}$ as shown in the equation \ref{eq112}.

\begin{equation} \label{eq112}
EDP_{system} = \eta ( EDP_{global} + EDP_{func} )
\end{equation}

Our objective function is shown in the equation \ref{eq91}. Our main objective is to minimize the system's EDP by choosing the optimal placement choice. 

\begin{equation} \label{eq91}
\textbf{Objective Function: Minimize }  EDP_{system}
\end{equation}

\subsection{Implementing Mapping Technique in MSP430FR6989} \label{MV}
Once we obtain the placement information from the $ILP\_solver$, we map the respective variables and the sections of a function to either SRAM or FRAM. We modify the linker script accordingly for mapping the sections or variables to either SRAM or FRAM. In our proposed mapping policy, placing global variables is straightforward, i.e., mapping the respective variable to either SRAM or FRAM based on the ILP decision.

We observed that from the linker script, we could map the whole stack section of each function to either SRAM or FRAM. We analyzed the mappings of the stack section for each function by modifying the linker script. We used the inbuilt attributes to differentiate mappings between SRAM and FRAM; for instance, we used the inbuilt attribute $( \_\_attribute\_\_((ramfunc)) $ that maps that function to SRAM. If we want to place the stack section to SRAM, we modify the linker script by replacing the default setting with " .stack: \{\} > RAM (HIGH) ". If we want to place the stack section to FRAM, we modify the linker script by replacing the default setting with " .stack: \{\} > FRAM". 

Similarly, for the text section, we observed that placing the text section in either SRAM or FRAM shows an impact on EDP. This effect is because the majority of access in the text section are read accesses, as we observed that the energy consumption for each read access to SRAM/FRAM differs. Table \ref{tab2} shows that approximately FRAM consumes 2x more read energy than SRAM. Thus, we analyzed each application where to map the text section based on the free space available. If we have enough space available in SRAM, we place the text section in SRAM itself; otherwise, we place the text section in FRAM. We included the following four lines in our linker script to check the above condition and map the text section.

\begin{enumerate}
     \item $ \#ifndef \_\_LARGE\_CODE\_MODEL\_\_$
  \item  .text             : \{\} > FRAM         
\item \#else
  \item  .text             : \{\} >> SRAM
\end{enumerate}
   
We modified the linker script for mapping the data section by using the inbuilt compiler directives. We followed the below three steps.  

\begin{enumerate}
    \item Allocate a new memory block, for instance, $NEW\_DATASECTION$. We can declare the start address and size of the data section in the linker script. 
    \item Define a segment (.Localvars) which stores in this memory block ($NEW\_DATASECTION$).
    \item Use \#pragma $DATA\_SECTION ( funct\_name, seg\_name)$ in the program to define functions in this segment. Where $funct\_name$ is the function name, and $seg\_name$ is the created segment name. For instance, \#pragma $DATA\_SECTION ( func\_1, .Localvars)$
\end{enumerate}

Once we are done with creating the different sections, we can allocate these sections to either SRAM or FRAM based on ILP decisions. For instance, placing " $NEW\_DATASECTION$: \{\} > FRAM" in the linker script, which maps the $NEW\_DATASECTION$ to FRAM.

\subsection{Support for Intermittent Computing} \label{IC}

When the power is stable, everything works properly. Because of the static allocation scheme, we map all functions/variables to SRAM/FRAM for the first time. During a power failure, SRAM and registers lose all of their contents, including mapping information. When power is restored, we don't know what functions/variables were allocated to SRAM before the failure. As a result, we must either restart the execution from the beginning or end up with incorrect results. Restarting the application consumes extra energy and time, making our system inefficient in terms of energy consumption and performance.

We propose a backup strategy during frequent power failures. FRAM was divided into $FRAM_n$ and $FRAM_b$ as shown in the figure \ref{md1}. $FRAM_n$ has a size of 125 KB and is used for regular mappings. $FRAM_b$ has a size of 3 KB that serves as a backup region (BR) during power failures. So, during a power failure, we back up all register and SRAM contents to FRAM. Whenever power is restored, we restore the register and SRAM contents from $FRAM_b$ to SRAM and resume the application execution. The proposed backup strategy reduces extra energy consumption and makes the system more energy efficient.

\subsubsection{Implementation details of Flash-based Programming for Intermittent Computing: }
MSP430F5529 consists of SRAM and Flash at main memory. SRAM is the only memory on the chip where the CPU can read code for executing the application during Flash programming. We need to copy the Flash program function onto the stack whenever we want to use only SRAM for mapping the application. Whenever we want to switch between SRAM to Flash, we need to restore the stack pointer and as well as we need to map the program counter register to the Flash memory region. 

During a power failure scenario, we must perform the backup operation to copy the SRAM data to the Flash memory region. For the backup operation, we made some changes to the inbuilt MSP430 functions, such as void Flash\_wb( char *Data\_ptr, char byte ) and void Flash\_ww( int *Data\_ptr, int word ). Where Flash\_wb() helps in writing the byte to the Flash memory region, Flash\_ww() helps in writing the word to the Flash memory region. 

Whenever power comes back, we must restore the contents from the Flash-based backup region to the SRAM memory region. We used the inbuilt functions, i.e., ctpl() functions for copying from Flash to SRAM, and after restoring, we needed to clear the Flash-based backup region; for this, we made changes to the inbuilt function, i.e., void Flash\_clr( int *Data\_ptr ) to clear the Flash data.


\section{Experimental Setup and Results} \label{exp1}
\subsection{Experimental Setup}
\begin{figure}[htbp]
\includegraphics[width= 0.8\linewidth]{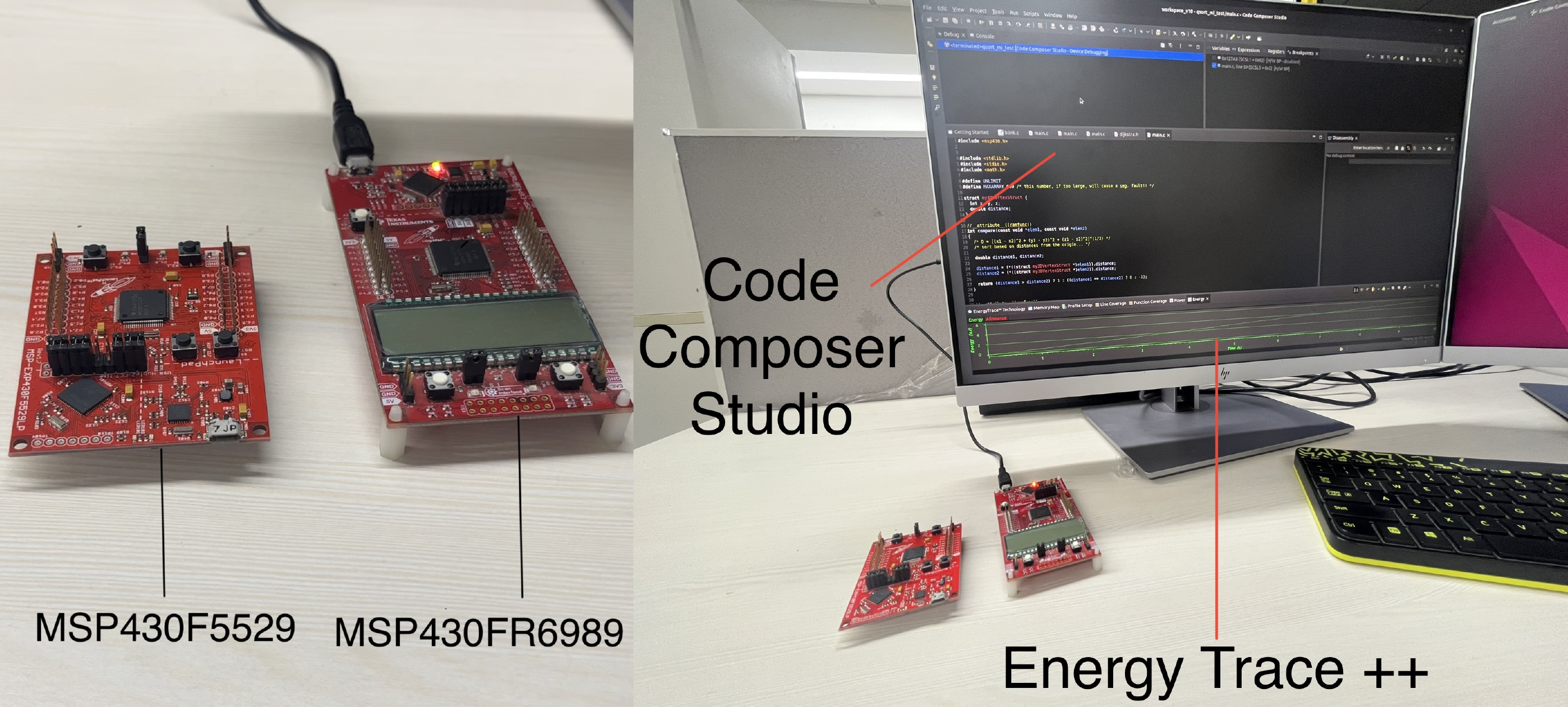}
\caption{(a) TI-based MSP430 Launchpad Development Boards (b) Working with EnergyTrace++ on CCS}
\label{exp}
\end{figure}

We used TI's MSP430FR6989 for all experiments. We experimented on mixed benchmarks, which have both Mi-Bench \cite{14} and TI-based benchmarks. We have shown the experimental setup in the table \ref{tab1}. The development platform and experimental setup are shown in figure \ref{exp}. We performed experiments to determine the energy required for a single read/write to SRAM/FRAM, as shown in the table \ref{tab2}. We collected the number of reads/writes for each global variable and functions as part of a one-time characterization. We also used TI's MSP430F5529 for comparing flash with FRAM. We performed experiments to determine the energy required for a single read/write to flash, as shown in the table \ref{tab2}.

\begin{table}[htp]
\centering
\caption{Experimental Setup}
\label{tab1}
\begin{tabular}{|l|l|}
\hline
\textbf{Component}        & \textbf{Description}             \\ \hline
\textbf{Target Board}     & TI MSP430FR6989 Launchpad        \\ \hline
\textbf{Core}             & MSP430 (1.8-3.6 V; 16 MHz)       \\ \hline
\textbf{Memory}           & 2KB SRAM and 128KB FRAM          \\ \hline
\textbf{IDE}              & Code Composer Studio             \\ \hline
\textbf{Energy Profiling} & Energy Trace++                   \\ \hline
\textbf{ILP Solver}       & LPSolve\_IDE                     \\ \hline
\textbf{Benchmarks}       & Mixed benchmarks (MiBench and TI-based)  \\ \hline
\end{tabular}%
\end{table}

MCU, which we experimented has MSP430 architecture, which is more suitable for IoT devices. The majority of MSP430 software is written in C and compiled with one of TI's recommended compilers ( IAR Embedded Code Bench, Code-Composer Studio (CCS), or msp430-gcc). The IAR Embedded Code Bench and CCS compilers are part of integrated development environments (IDEs). We used the widely used, freely available, and easily extended tool, i.e., CCS, for all experiments in this article. EnergyTrace++ technology allows us to calculate energy and power consumption directly. According to the datasheet for the MSP430FR6989, the number of cycles required to read/write in FRAM is twice that of SRAM, which means the access penalty of FRAM is twice that of SRAM at this specific operating point of 16 MHz. The latency penalty disappears when operating at/below 8 MHz and gets worse above 16 MHz. 



\begin{table*}[htp]
\centering
\caption{Energy Values for each read/write to SRAM and FRAM}
\label{tab2}
\begin{tabular}{|l|l|l|}
\hline
\textbf{Memory} & \textbf{Per Read Energy (nJ)} & \textbf{Per Write Energy (nJ)} \\ \hline
\textbf{SRAM}   &               5500                &        5600                        \\ \hline
\textbf{FRAM}   &        10325                       &        13125                        \\ \hline
\textbf{Flash}   &        23876                       &        31198                       \\ \hline
\end{tabular}%
\end{table*}

\subsection{Evaluation Benchmarks} \label{p511}
We chose benchmarks from both the MiBench suite and TI benchmarks. One of the primary motivations for using the MiBench suite is that most of the TI-based benchmarks were small in size and easily fit into either SRAM or FRAM. In these cases, we don't require any hybrid memory design. Most of the TI-based benchmarks have only one or two functions and 3-4 global variables, which is not useful for the hybrid main-memory design. Thus we used mixed benchmarks consisting of 4 TI-based benchmarks and 12 from the MiBench suite.

For the MiBench suite, we first make MCU-compatible benchmarks by adding MCU-related header files and watchdog timers. All benchmarks may not be compatible with the MCU. Thus, we need to choose the benchmarks from the MiBench suite, which are compatible with the MSP430 boards. Once we have benchmarks, we execute them on board for the machine code. Using the .asm file, we calculate the inputs that are required by the ILP solver, as shown in figure \ref{f1}.

\subsection{Baseline Configurations} \label{p51}
We chose five different memory configurations to compare with the proposed memory configuration. 

We directly map all the functions/variables to FRAM in the FRAM-only configuration, as shown in figure \ref{b11}. We use FRAM-only configuration to compare our proposed memory configuration during stable and unstable power scenarios. 

We directly map all the functions/variables to SRAM in the SRAM-only configuration, as shown in figure \ref{b11}. We use SRAM-only configuration to compare our proposed memory configuration during stable and unstable power scenarios.

We used the empirical method of Jayakumar et al. \cite{backup}. We compare this configuration with our proposed configuration during stable and unstable power scenarios to observe the importance of the proposed than the existing work.

In the SRAM+Flash with ILP configuration, we used the proposed ILP technique for the flash-based msp430 board \cite{msp1}. We compare this configuration with our proposed configuration during stable and unstable power scenarios to observe the difference between FRAM and Flash technologies.

In the SRAM+FRAM with ILP configuration, we have the proposed memory mapping technique that doesn't support BR. We compare this configuration with our proposed configuration during frequent power failures to observe the importance of BR. The overview of all baseline configurations is shown in table \ref{base}. The experimental setup for all the above five configurations is the same as the one proposed.

\begin{table}[htp]
\centering
\caption{Overview of the Different Memory Configurations for Comparing with the Proposed Memory Configuration}
\label{base}
\begin{tabular}{|llllll|}
\hline
\multicolumn{1}{|l|}{\textbf{Configuration}} & \multicolumn{1}{l|}{\textbf{FRAM}} & \multicolumn{1}{l|}{\textbf{SRAM}} & \multicolumn{1}{l|}{\textbf{Flash}} & \multicolumn{1}{l|}{\textbf{Backup Region (BR)}} & \textbf{ILP} \\ \hline
\multicolumn{1}{|l|}{FRAM-only}             & \multicolumn{1}{l|}{\cmark}              & \multicolumn{1}{l|}{\xmark}              & \multicolumn{1}{l|}{\xmark}               & \multicolumn{1}{l|}{\xmark}                            &  \xmark            \\ \hline
\multicolumn{1}{|l|}{SRAM-only}             & \multicolumn{1}{l|}{\xmark}              & \multicolumn{1}{l|}{\cmark}              & \multicolumn{1}{l|}{\xmark}               & \multicolumn{1}{l|}{\xmark}                            &     \xmark         \\ \hline
\multicolumn{1}{|l|}{Jayakumar et al. \cite{backup}}             & \multicolumn{1}{l|}{\cmark}              & \multicolumn{1}{l|}{\cmark}              & \multicolumn{1}{l|}{\xmark}               & \multicolumn{1}{l|}{\xmark}                            &   \xmark           \\ \hline
\multicolumn{1}{|l|}{SRAM+Flash with ILP}             & \multicolumn{1}{l|}{\xmark}              & \multicolumn{1}{l|}{\cmark}              & \multicolumn{1}{l|}{\cmark}               & \multicolumn{1}{l|}{\xmark}                            &  \cmark          \\ \hline
\multicolumn{1}{|l|}{SRAM+FRAM with ILP}             & \multicolumn{1}{l|}{\cmark}              & \multicolumn{1}{l|}{\cmark}              & \multicolumn{1}{l|}{\xmark}               & \multicolumn{1}{l|}{\xmark}                            &    \cmark          \\ \hline

\multicolumn{1}{|l|}{\textbf{Proposed}}               & \multicolumn{1}{l|}{\cmark}              & \multicolumn{1}{l|}{\cmark}              & \multicolumn{1}{l|}{\cmark}               & \multicolumn{1}{l|}{\cmark}                            & \cmark             \\ \hline

\multicolumn{6}{|l|}{\cmark - Supported ,   \xmark - Not Supported}                                                                                                                                                                                                         \\ \hline

\end{tabular}%
\end{table}

\subsection{Results}
The proposed memory configuration is evaluated in this section under stable and unstable power. The proposed memory configuration is compared with five different memory configurations as discussed in the section \ref{p51}.

\subsubsection{\textbf{Under Stable Power}:}

\begin{figure}[htbp]
\includegraphics[width= 0.8\linewidth]{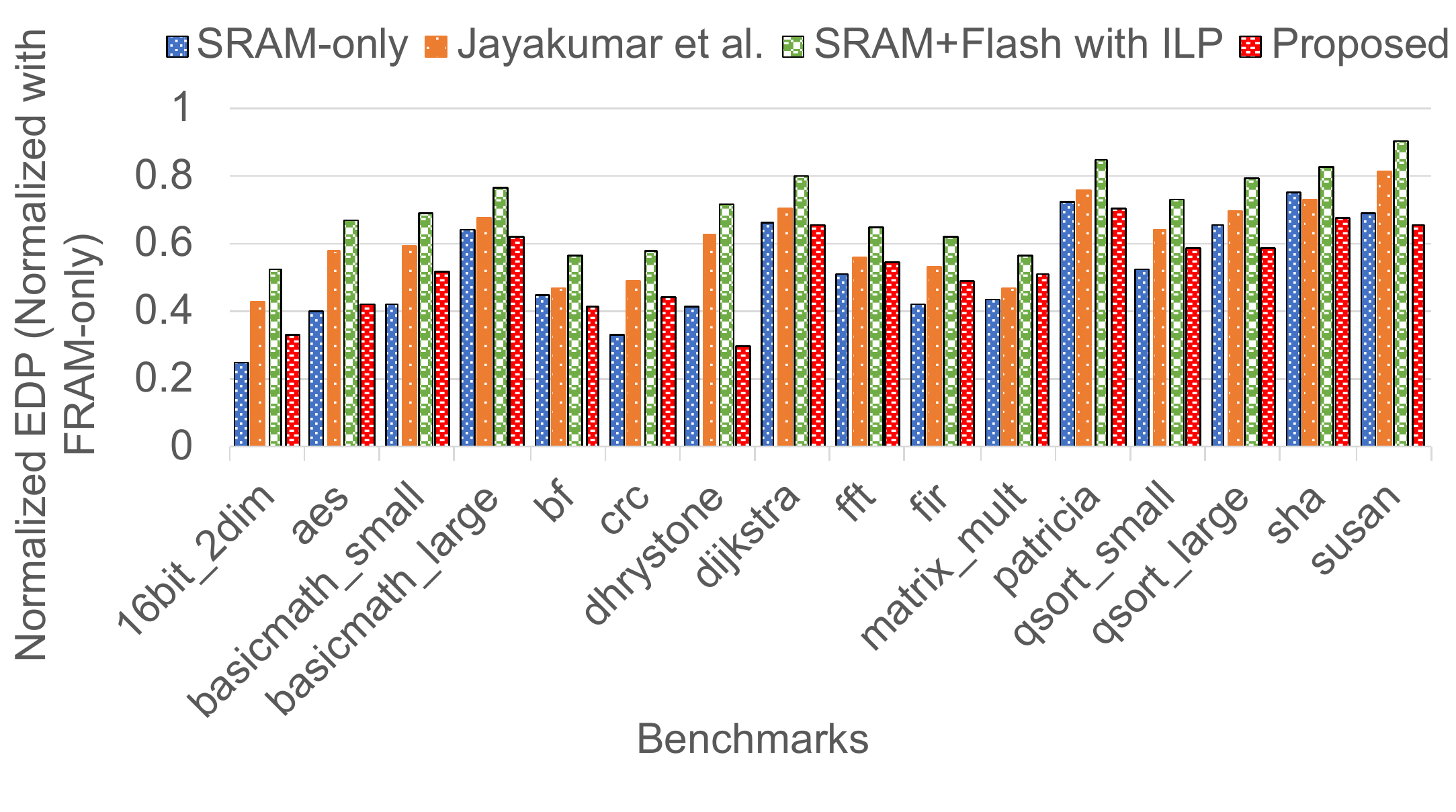}
\caption{Comparison between Baseline configurations and the Proposed under Stable Power}
\label{fig2a}
\end{figure}

Our main objective of the proposed memory configuration is to minimize the system's EDP. All values shown in figure \ref{fig2a} are normalized with the FRAM-only configuration. Compared to the FRAM-only configuration, the proposed gets 38.10\% lesser EDP, as shown in figure \ref{fig2a}. Because there are no power interruptions in this scenario, this improvement is totally from the proposed ILP model. In configuration-1, we place everything to FRAM, where FRAM consumes more energy and the number of cycles than SRAM, as shown in the table \ref{tab2}. Our proposed memory configuration incorporates the placement recommendation from the proposed ILP model and suggests utilizing both SRAM and FRAM. 

Under a stable power scenario, the proposed gets 9.30\% less EDP than Jayakumar et al., as shown in figure \ref{fig2a}. We discussed the author's empirical model and assumptions in the previous section \ref{moti}. The authors assumed that the data section included all global variables, constants, and arrays. As a result, our proposed ILP-based mapping differs from the author's mapping in that our proposed mapping outperforms the existing work. Under stable power, Jayakumar et al. receive 24.57\% less EDP than the FRAM-only configuration, as shown in figure \ref{fig2a}. This advantage is primarily due to Jayakumar et al. hybrid memory.

In comparison to SRAM+Flash with ILP configuration, the proposed reduces EDP by 18.55\%, as shown in figure \ref{fig2a}. We used flash+SRAM with our proposed ILP framework in this configuration. As shown in table \ref{tab2}, the above benefit is primarily due to FRAM because flash consumes more energy. Jayakumar et al. outperform SRAM+Flash with ILP configuration during stable power. Because of FRAM in Jayakumar et al., even our proposed ILP model is ineffective in this case. We encountered that Jayakumar et al. achieved 9.19\% less EDP than SRAM+Flash with ILP configuration, and this benefit is because of smaller applications. From figure \ref{fig2a}, SRAM+Flash with ILP configuration performs better for large applications than SRAM+Flash with ILP configuration. Jayakumar et al. empirical method suggest placing more content on SRAM because SRAM is sufficient for placing the entire small-size application. As a result, the performance of Jayakumar et al. is dependent on the application size, as for large-size applications, even FRAM does not outperform Flash.

Comparing the proposed memory configuration to the empirical method of Jayakumar et al. helps in understanding the role of the ILP model. This comparison also clarifies whether these advantages stem from the mapping granularity or the ILP. The proposed memory configuration outperforms the existing one, demonstrating that it benefits from mapping granularity and the ILP model.

SRAM-only configuration outperforms the proposed and all other memory configurations under stable power conditions. We noticed that this benefit is primarily due to SRAM, but it only applies to smaller applications. SRAM-only achieves 36.19\% less EDP than the proposed for smaller applications, as shown in figure \ref{fig2a}. We also looked at large applications where the proposed outperforms the SRAM-only configuration by a small margin. When the SRAM is full, the MCU must wait for the space to be released, which consumes extra energy and cycles. For more extensive applications, SRAM-only configuration achieves 2.94\% more EDP than proposed.

\begin{figure}[htbp]
\includegraphics[width= 0.8\linewidth]{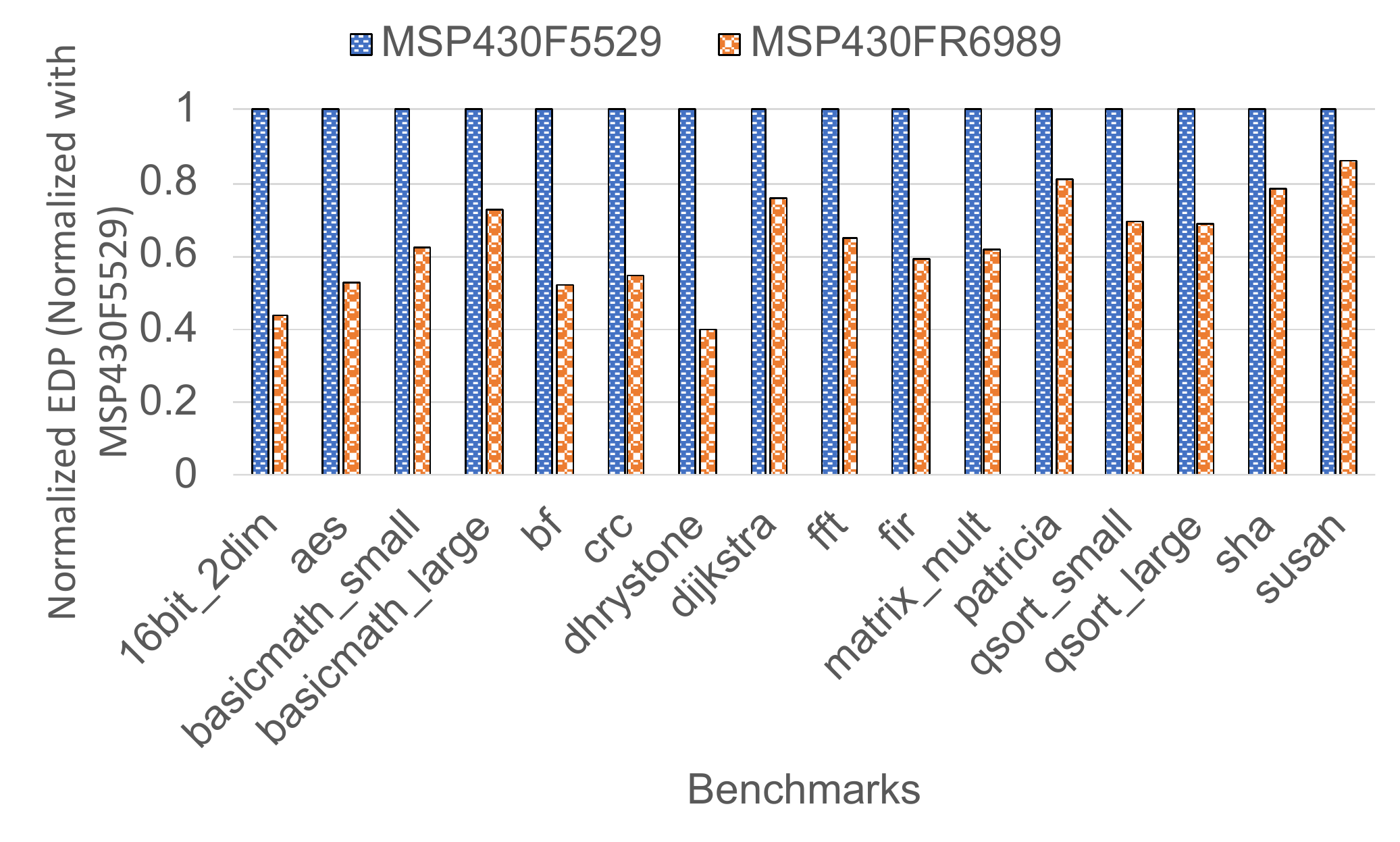}
\caption{omparison between MSP430FR6989 (FRAM-based MCU) and MSP430F5529 (Flash-based MCU) under Stable Power}
\label{fig2a1}
\end{figure}

We also evaluated our proposed framework with another MSP430F5529 MCU with flash and SRAM for completeness. This comparison assists the user in selecting the most appropriate NVM technology, such as FRAM or Flash, as needed. To be fair, we used the same sizes of SRAM (2 KB) and Flash (128 KB) in this comparison. We compared FRAM-based and flash-based MCUs under stable power conditions. We used the proposed frameworks and techniques in both MCUs. We discovered that the proposed FRAM-based configuration outperforms the flash-based configuration. Flash-based configurations consume 26.03\% more EDP than FRAM-based configurations, as shown in figure \ref{fig2a1}. Flash consumes more energy, as shown in table \ref{tab2}. 

\subsubsection{\textbf{Under Unstable power}:}
We used the default TI-based compute through power loss (ctpl) tool for migration. During a power failure, we need to migrate the SRAM contents to a FRAM-based backup region ($FRAM_b$), i.e., the backup process. Whenever power comes back, we need to migrate the ($FRAM_b$) contents to SRAM, i.e., the restoration process. So, all these migrations are done using ctpl() functions. We introduce a power failure by changing the low power modes mentioned in the MSP430FR6989 design document. We used ctpl() for creating power failures. We assume that the number of power failures is spread equally within the execution period. For instance, if the total execution period for an application is 20 milliseconds (ms), and let's say the number of power failures is four, then for every 5 ms, we experience a power failure. 

Considering energy harvesting sources, such as piezoelectric and vibration-based sources, they extract significantly less energy from their surroundings. In such cases, the capacitor is unable to store sufficient energy, leading to frequent power failures. As a consequence, our proposed architecture is capable of handling these worst-case scenarios. However, existing works by Xie et al. \cite{xie} and Badri et al. \cite{m7,m8} made similar assumptions that almost every power failure occurs every 200 and 500 ms for a 1-core CPU running at a frequency of 480 MHz.

\begin{figure}[htbp]
\includegraphics[width= 0.8\linewidth]{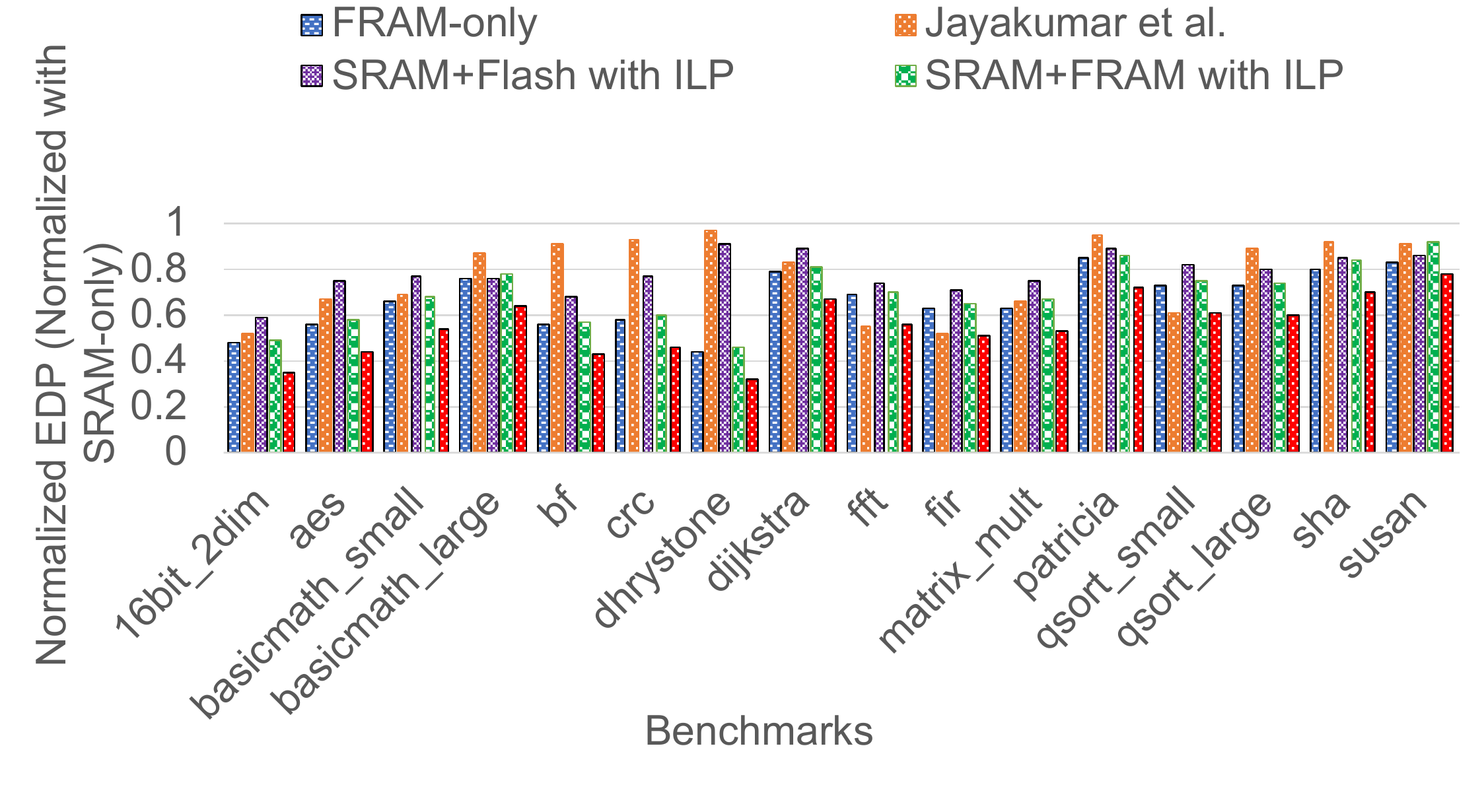}
\caption{Comparison between Baseline configurations and the Proposed under Unstable Power}
\label{fig3a}
\end{figure}

We performed experiments under unstable power to compare the proposed memory configuration with five memory configurations, as shown in table \ref{base}. All values shown in figure \ref{fig3a} are normalized with SRAM-only configuration. Compared to the SRAM-only configuration, the proposed gets 15.97\% lesser EDP, as shown in figure \ref{fig3a}. We observed that migration overhead is less than the energy consumed to execute the application from FRAM, and this migration overhead depends on the number of power failures. For instance, one backup migration consumes approximately 16.88 mJ of energy, and one restore migration consumes approximately 11.606 mJ of energy in a qsort application. The above benefit to our proposed configuration is using a hybrid memory. 

Under an unstable power scenario, the proposed gets 21.99\% less EDP than Jayakumar et al., as shown in figure \ref{fig3a}. We discussed the author's empirical model and assumptions in the previous section \ref{moti}. As already stated, the Jayakumar et al. empirical method is more beneficial for small applications. In contrast, the author's empirical method suggests placing more content on SRAM because SRAM is sufficient for placing the entire small-size application. Thus, for [20] work, backup/restore operations take more energy during a power failure. Our proposed mapping outperforms the existing work. During frequent power failures, Jayakumar et al. receive 6.91\% less EDP than the FRAM-only configuration, as shown in figure \ref{fig3a}. This advantage is primarily due to Jayakumar et al. hybrid memory. 

Compared to SRAM+Flash with ILP configuration, the proposed reduces EDP by 23.05\%, as shown in figure \ref{fig3a}. As shown in table \ref{tab2}, the above benefit is primarily due to FRAM because flash consumes more energy. Jayakumar et al. outperform SRAM+Flash with ILP configuration during stable power. Because of FRAM in Jayakumar et al., even our proposed ILP model is ineffective for this comparison. We encountered that Jayakumar et al. achieved 6.28\% less EDP than SRAM+Flash with ILP configuration for smaller applications. The above benefit for Jayakumar et al. is minimal because the size of backup/restores increases, which even neutralizes the flash for some applications, as shown in figure \ref{fig3a}. SRAM+Flash with ILP configuration achieves 2.69\% less EDP than Jayakumar et al. for large applications, as shown in figure \ref{fig3a}. As a result, the performance of Jayakumar et al. is dependent on the application size, as for large-size applications, even FRAM does not outperform Flash.

The proposed outperforms all memory configurations under unstable power conditions. This benefit is primarily due to a hybrid memory and the proposed mapping technique. SRAM-only achieves 42.98\% less EDP than the proposed, as shown in figure \ref{fig3a}.

When we remove BR, all the mapping information of SRAM is lost because our model is static. We introduce a BR in the FRAM memory region to save this mapping information. During a power failure, we migrate the SRAM contents to $FRAM_b$; whenever power comes back, we restore the $FRAM_b$ contents to the SRAM.

We experimented to know the importance of BR, where we compared the proposed memory configuration with SRAM+FRAM with ILP configuration. Compared to SRAM+FRAM with ILP configuration, the proposed gets 23.94\% lesser EDP, as shown in figure \ref{fig3a}. This benefit is because we need to re-execute the application four times from the beginning, which consumes extra time and energy. The number of times re-executing the application is equal to the number of power failures.

\begin{figure}[htbp]
\includegraphics[width= 0.8\linewidth]{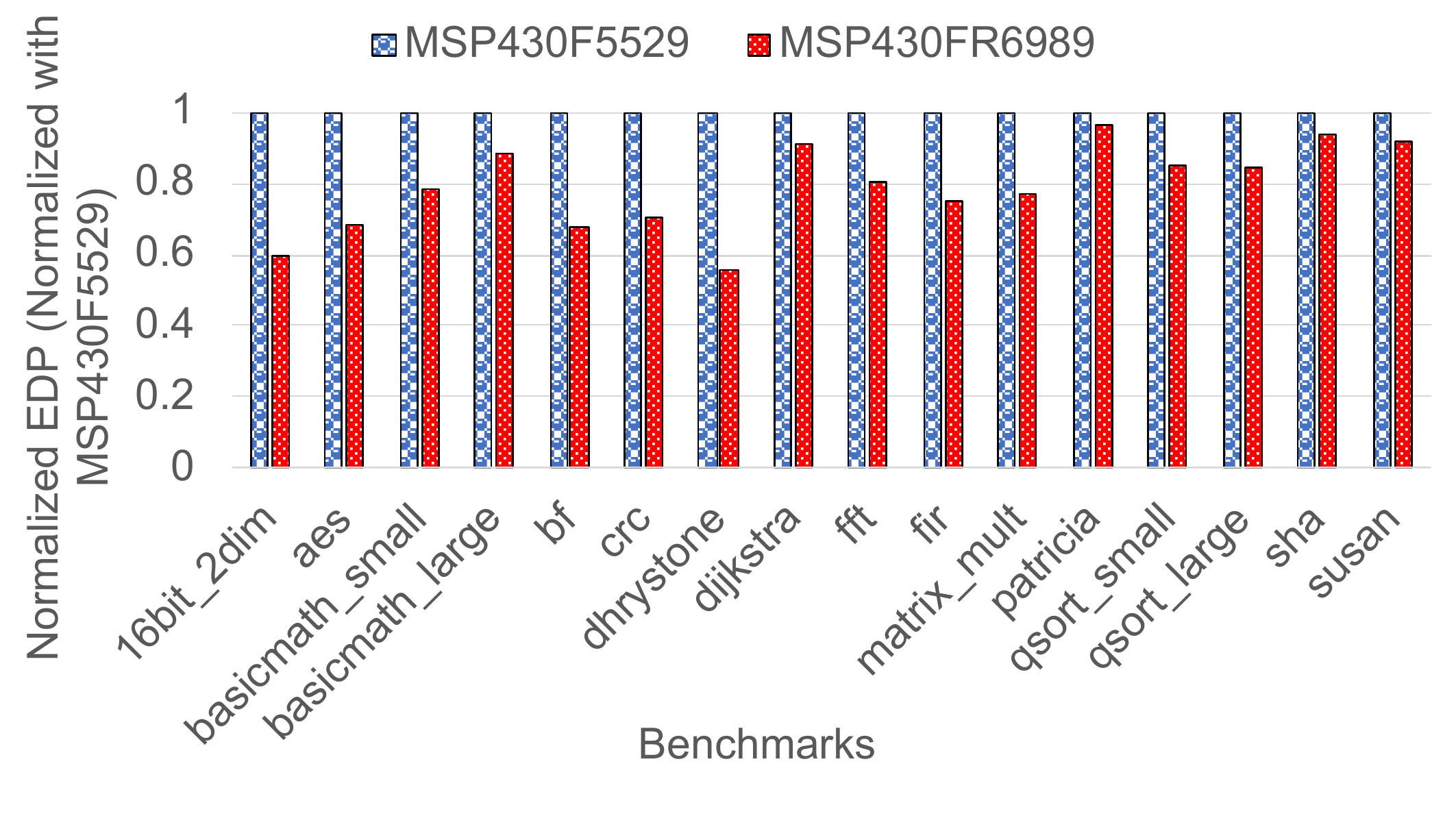}
\caption{Comparison between MSP430FR6989 (FRAM-based MCU) and MSP430F5529 (Flash-based MCU) under Unstable Power}
\label{fig3a1}
\end{figure}

We also evaluated our proposed framework with another MSP430F5529 MCU, which consists of flash and SRAM for completeness. This comparison assists the user in selecting the most appropriate NVM technology, such as FRAM or Flash, as needed. To be fair, we used the same sizes of SRAM (2 KB) and Flash (128 KB) in this comparison. We also used BR in these experiments; the only difference is that we replaced FRAM with the flash in the proposed configurations, and everything is the same. We compared FRAM-based and flash-based MCUs under unstable power conditions. We used the proposed frameworks and techniques in both MCUs. We discovered that the proposed FRAM-based configuration outperforms the flash-based configuration. Flash-based configurations consume 16.50\% more EDP than FRAM-based configurations, as shown in figure \ref{fig3a1}. Flash consumes more energy, as shown in table \ref{tab2}. 

\subsection{Summary of the Proposed Mapping Technique}

We outline the proposed ILP-based memory mapping technique in this section. Following all of these analyses, we observed that the mappings shown below consume less EDP than other design choices, as shown in the table. To keep things simple, we only showed the final mapping configurations for each application's stack, data, and text sections, keeping out the final mappings for global variables.

\begin{table}[htp]
\caption{Optimal Placement for different Applications in MSP430FR6989}
\label{tab5}
\begin{tabular}{|l|l|l|l|}
\hline
\textbf{Benchmarks} & \textbf{Stack} & \textbf{Text} & \textbf{Data} \\ \hline
16bit\_2dim         & SRAM           & SRAM          & SRAM          \\ \hline
aes                 & SRAM           & FRAM          & FRAM          \\ \hline
basicmath\_small    &  SRAM              &  SRAM             & FRAM              \\ \hline
basicmath\_large    &   SRAM             & FRAM              &  FRAM             \\ \hline
bf                  &  SRAM              & SRAM              &  FRAM             \\ \hline
crc                 &  SRAM              &  FRAM             &  SRAM             \\ \hline
dhrystone           &  FRAM              &  SRAM             &  FRAM             \\ \hline
dijkstra            &  SRAM              &  FRAM             &    SRAM           \\ \hline
fft                 &  SRAM              &   SRAM            &    FRAM           \\ \hline
fir                 &  SRAM              &  SRAM             &   FRAM            \\ \hline
matrix\_mult        &  SRAM              &   SRAM            &   SRAM            \\ \hline
patricia            &   SRAM             &   FRAM            &  SRAM             \\ \hline
qsort\_small        &  SRAM              &    SRAM           &  FRAM             \\ \hline
qsort\_large        &  SRAM              &   FRAM            &  FRAM             \\ \hline
sha                 &  SRAM              &   FRAM            &  FRAM             \\ \hline
susan               &   SRAM             &     FRAM          &  FRAM             \\ \hline
\end{tabular}
\end{table}

Table \ref{tab5} shows that, with the exception of the dhrystone application, the remaining three TI benchmark applications (fir, matrix, and 16bit\_2dim) are very small and can easily be placed in SRAM. We don't need FRAM for these types of smaller applications, but there is a disadvantage during frequent power failures. Backup and restore sizes to FRAM are larger for these applications during frequent power failures. As a result, our proposed backup/restore strategy should be intelligent enough to reduce EDP. The dhrystone application, on the other hand, has a larger stack section that requires FRAM to accommodate the entire stack section.

As we can see from the table \ref{tab5}, many applications used both SRAM and FRAM for the Mi-Bench applications. As a result, we can conclude that a hybrid main memory design is required for many applications. Using a hybrid main memory design helps to reduce EDP during stable power scenarios. Even so, determining how and where to backup the volatile contents can be difficult during frequent power outages. However, our proposed memory mapping technique and the framework suggest using a hybrid main memory design that supports intermittent computing.

\section{Conclusions} \label{p6}
This paper proposed an ILP-based memory mapping technique that reduces the system's energy-delay product. For both global variables and functions, we formulated an ILP model. Functions consist of data, stack, and code sections. Our ILP model suggests placing each section on either SRAM or FRAM. Under both stable and unstable power scenarios, we compared the proposed memory configuration to the baseline memory configurations. We evaluated our proposed frameworks and techniques on actual boards. We added a backup region in FRAM to support intermittent computing. We compared the proposed framework with the recent related work. 

Under stable power, our proposed memory configuration consumes 38.10\% less EDP than the FRAM-only configuration and 9.30\% less EDP than the existing work. Under unstable power, our proposed configuration achieves 15.97\% less EDP than the FRAM-only configuration and 21.99\% less EDP than the existing work. Under stable power, our proposed memory configuration consumes 18.55\% less EDP than SRAM+Flash with ILP configuration. We also compared FRAM-based MSP430FR6989 with flash-based MSP430F5529. Compared to the flash, the FRAM-based hybrid main memory design consumes less EDP. FRAM-based design consumes 26.03\% less EDP than flash-based design during stable power and 16.50\% less EDP than flash-based during frequent power failures. 

\section*{Acknowledgement}
This work is supported by the grant received from the Department of Science and Technology (DST), Govt. of India, for the Technology Innovation Hub at the IIT Ropar in the framework of the National Mission on Interdisciplinary Cyber-Physical Systems.
%
\bibliographystyle{ACM-Reference-Format}
\bibliography{main}

\end{document}